\documentclass[aps,prl,twocolumn,superscriptaddress,showpacs,showkeys,amsmath,amssymb,floatfix]{revtex4}
\usepackage{booktabs,graphicx,mathrsfs,verbatim}
\usepackage[colorlinks=true,citecolor=blue,filecolor=blue,linkcolor=blue,urlcolor=blue,pdftex]{hyperref}
\usepackage[usenames,dvipsnames]{color}
\usepackage{ulem}
\usepackage{color}
\usepackage{setspace}
\usepackage{pdfpages}

\newcommand{\1}[1]{\, \mathrm{#1}} 
\newcommand{\n}[1]{\mathrm{#1}} 

\newcommand{\arxiv}[1]{\href{http://arxiv.org/abs/#1}{\texttt{arXiv:#1}}}

\newcommand{\assergi}{\affiliation{INFN, Laboratori Nazionali del Gran Sasso, Assergi, 67100, Italy}}
\newcommand{\bologna}{\affiliation{University of Bologna and INFN-Bologna, Bologna, Italy}}
\newcommand{\columbia}{\affiliation{Physics Department, Columbia University, New York, NY 10027, USA}}
\newcommand{\coimbra}{\affiliation{Department of Physics, University of Coimbra, R. Larga, 3004-516, Coimbra, Portugal}}
\newcommand{\heidelberg}{\affiliation{Max-Planck-Institut f\"ur Kernphysik, Saupfercheckweg 1, 69117 Heidelberg, Germany}}
\newcommand{\houston}{\affiliation{Department of Physics and Astronomy, Rice University, Houston, TX 77005 - 1892, USA}}

\newcommand{\losangeles}{\affiliation{Physics \& Astronomy Department, University of California, Los Angeles, USA}}
\newcommand{\mainz}{\affiliation{Institut f\"ur Physik, Johannes Gutenberg Universit\"at Mainz, 55099 Mainz, Germany}}
\newcommand{\munster}{\affiliation{Institut f\"ur Kernphysik, Wilhelms-Universit\"at M\"unster, 48149 M\"unster, Germany}}
\newcommand{\nikhef}{\affiliation{Nikhef  and the University of Amsterdam, Science Park, Amsterdam, Netherlands}}
\newcommand{\purdue}{\affiliation{Department of Physics, Purdue University, West Lafayette, IN 47907, USA}}
\newcommand{\shanghai}{\affiliation{Department of Physics, Shanghai Jiao Tong University, Shanghai, 200240, China}}
\newcommand{\subatech}{\affiliation{SUBATECH, Ecole des Mines de Nantes, CNRS/In2p3, Universit\'e de Nantes, 44307 Nantes, France}}
\newcommand{\torino}{\affiliation{University of Torino and INFN-Torino, Torino, Italy}}
\newcommand{\weizmann}{\affiliation{Department of Particle Physics and Astrophysics, Weizmann Institute of Science, 76100 Rehovot, Israel}}
\newcommand{\zurich}{\affiliation{Physics Institute, University of Z\"{u}rich, Winterthurerstr. 190, CH-8057, Switzerland}}

\begin{document}

\title{Dark Matter Results from 225\,Live Days of XENON100 Data}

\author{E.~Aprile}\columbia 
\author{M.~Alfonsi}\nikhef
\author{K.~Arisaka}\losangeles
\author{F.~Arneodo}\assergi
\author{C.~Balan}\coimbra
\author{L.~Baudis}\zurich
\author{B.~Bauermeister}\mainz
\author{A.~Behrens}\zurich
\author{P.~Beltrame}\losangeles
\author{K.~Bokeloh}\munster
\author{E.~Brown}\munster
\author{G.~Bruno}\assergi
\author{R.~Budnik}\columbia 
\author{J.~M.~R.~Cardoso}\coimbra
\author{W.-T.~Chen}\subatech
\author{B.~Choi}\columbia
\author{D.~Cline}\losangeles
\author{A.~P.~Colijn}\nikhef
\author{H.~Contreras}\columbia
\author{J.~P.~Cussonneau}\subatech
\author{M.~P.~Decowski}\nikhef
\author{E.~Duchovni}\weizmann
\author{S.~Fattori}\mainz
\author{A.~D.~Ferella}\zurich
\author{W.~Fulgione}\torino
\author{F.~Gao}\shanghai
\author{M.~Garbini}\bologna
\author{C.~Ghag}\losangeles
\author{K.-L.~Giboni}\columbia
\author{L.~W.~Goetzke}\columbia
\author{C.~Grignon}\mainz
\author{E.~Gross}\weizmann
\author{W.~Hampel}\heidelberg
\author{F.~Kaether}\heidelberg
\author{A.~Kish}\zurich
\author{J.~Lamblin}\subatech
\author{H.~Landsman}\weizmann
\author{R.~F.~Lang}\purdue\columbia
\author{M.~Le~Calloch}\subatech
\author{C.~Levy}\munster
\author{K.~E.~Lim}\columbia
\author{Q.~Lin}\shanghai
\author{S.~Lindemann}\heidelberg
\author{M.~Lindner}\heidelberg
\author{J.~A.~M.~Lopes}\coimbra
\author{K.~Lung}\losangeles
\author{T.~Marrod\'an~Undagoitia}\zurich
\author{F.~V.~Massoli}\bologna
\author{A.~J.~Melgarejo~Fernandez} \email{ajmelgarejo@astro.columbia.edu} \columbia
\author{Y.~Meng}\losangeles
\author{A.~Molinario}\torino
\author{E.~Nativ}\weizmann
\author{K.~Ni}\shanghai
\author{U.~Oberlack}\mainz\houston
\author{S.~E.~A.~Orrigo}\coimbra
\author{E.~Pantic}\losangeles
\author{R.~Persiani}\bologna
\author{G.~Plante}\columbia
\author{N.~Priel}\weizmann
\author{A.~Rizzo}\columbia
\author{S.~Rosendahl}\munster
\author{J.~M.~F.~dos Santos}\coimbra
\author{G.~Sartorelli}\bologna
\author{J.~Schreiner}\heidelberg
\author{M.~Schumann}\email{marc.schumann@physik.uzh.ch} \zurich
\author{L.~Scotto~Lavina}\subatech
\author{P.~R.~Scovell}\losangeles
\author{M.~Selvi}\bologna
\author{P.~Shagin}\houston
\author{H.~Simgen}\heidelberg
\author{A.~Teymourian}\losangeles
\author{D.~Thers}\subatech
\author{O.~Vitells}\weizmann
\author{H.~Wang}\losangeles
\author{M.~Weber}\heidelberg
\author{C.~Weinheimer}\munster

\collaboration{The XENON100 Collaboration}\noaffiliation

\begin{abstract}
We report on a search for particle dark matter with the XENON100 experiment, operated at the Laboratori Nazionali del Gran Sasso (LNGS) for 13~months during 2011 and 2012. XENON100 features an ultra-low electromagnetic background of $(5.3 \pm 0.6) \times 10^{-3}$\,events/(keV$_{\n{ee}}\times$kg$\times$day) in the energy region of interest. A blind analysis of 224.6\,live days $\times$ 34\,kg exposure has yielded no evidence for dark matter interactions. The two candidate events observed in the pre-defined nuclear recoil energy range of 6.6-30.5\,keV$_{\n{nr}}$ are consistent with the background expectation of $(1.0 \pm 0.2)$\,events. A Profile Likelihood analysis using a 6.6-43.3\,keV$_{\n{nr}}$ energy range sets the most stringent limit on the spin-independent elastic WIMP-nucleon scattering cross section for WIMP masses above 8\,GeV/$c^2$, with a minimum of $2 \times 10^{-45}$\,cm$^2$ at 55\,GeV/$c^2$ and 90\% confidence level.
\end{abstract}

\pacs{
 95.35.+d, 
 14.80.Ly, 
 29.40.-n, 
}

\keywords{Dark Matter, Direct Detection, Xenon}

\maketitle

Astronomical and cosmological observations indicate that a large amount of the energy content of the Universe is made of dark matter~\cite{Jarosik:2010iu;Nakamura:2010zzi}. Particle candidates under the generic name of Weakly Interacting Massive Particles (WIMPs)~\cite{Steigman:1984ac;Jungman:1995df} arise naturally in many theories beyond the Standard Model of particle physics, such as supersymmetry, universal extra dimensions, or little Higgs models. The search for these particles continues with a variety of experimental approaches~\cite{Bertone:2005}. 
In direct detection experiments, one attempts to observe the nuclear recoils~(NRs) produced by WIMP scattering off nucleons~\cite{Goodman:1984dc}. The recoil spectrum falls exponentially with energy and extends to a few tens of~keV only. The expected low event rate requires large detectors built from radio-pure materials and that are capable of identifying and rejecting backgrounds from various sources. 

The XENON100 experiment, described in detail in~\cite{Aprile:2012instr}, uses liquid xenon (LXe) as both WIMP target and detection medium, with simultaneous measurement of the ionization and scintillation signals produced by particle interactions in the active volume. The detector is a cylindrical two-phase (gas/liquid) time projection chamber (TPC) with a LXe target mass of 62\,kg. An additional 99\,kg of the same high-purity LXe, optically separated from the target volume, is instrumented as an active scintillator veto. The TPC and the veto are mounted in a double-walled stainless-steel cryostat, enclosed by a passive shield made from OFHC copper, polyethylene, lead, and water/polyethylene. The shield is continuously purged with boil-off~N$_2$ gas in order to suppress radon backgrounds. The LXe is kept at the operating temperature of about $-91^\circ$C by a pulse tube refrigerator (PTR) mounted outside the shield. For the run leading to this new result the PTR has been in continuous operation for a total of $\sim$20\,months. This is the first demonstration, to our knowledge, of a LXe detector operated over such a long period of time. 

The key feature of the XENON100 TPC is its ability to reconstruct the energy and three-dimensional coordinates on an event-by-event basis. This enables background reduction by fiducial volume optimization, exploiting the self-shielding of LXe. An energy deposition in the TPC produces both ionization electrons and scintillation photons. The electrons, moved from the interaction point by a drift field of 530\,V/cm, are extracted from the liquid and accelerated in the gas by a $\sim$12\,kV/cm field, producing proportional scintillation light. The amplified charge signal~(S2) and the direct scintillation signal~(S1) are both detected by two arrays of 1''-square Hamamatsu R8520-AL photomultipliers (PMTs), selected for low radioactivity and high quantum efficiency~\cite{Aprile:2012instr}. One array is immersed in the LXe below the cathode of the TPC for optimal light collection, and one is placed in the xenon gas above the amplification gap.

The $z$-position of a particle interaction in the TPC is reconstructed, with a precision of $0.3\1{mm}$ ($1\sigma$), from the time difference between the S1 and S2~signals and the known electron drift velocity. 
The localized distribution of the S2 signal over the top PMTs is used to obtain the ($x,y$)-coordinates using a Neural Network algorithm with an uncertainty $<3$\,mm ($1\sigma$)~\cite{Aprile:2012instr,Aprile:2011run08}. 
The spatial reconstruction also allows for the rejection of multiple-scatter events, such as from neutrons, since WIMPs are expected to interact only once. Double-scatter events can be separated when their vertices differ by $\Delta z>3$\,mm. Finally, the ratio S2/S1 is different for NRs (WIMPs, neutrons) and electronic recoils (ER; $\beta$/$\gamma$-background) and is also used for background discrimination. 

XENON100 is installed at the Laboratori Nazionali del Gran Sasso (LNGS) of INFN, Italy, at an average depth of 3600\,m water equivalent, where the muon flux is suppressed by $10^6$ with respect to sea level~\cite{ref::lvd}. Due to careful material selection~\cite{Aprile:2011ru} and detector design~\cite{Aprile:2012instr}, the total ER background of XENON100 in an inner 34~kg fiducial volume is  $5.3\times10^{-3}$\,events/(keV$_{\n{ee}}\times$kg$\times$day) (keV$_{\n{ee}}=$ keV electron-equivalent~\cite{Aprile:2008rc}) in the dark matter energy region, before S2/S1 discrimination.

Compared to the results reported in~\cite{Aprile:2011run08}, the new dark matter search is characterized by a considerably larger exposure and a reduction of a factor 20 of the intrinsic back\-ground from $^{85}$Kr, by cryogenic distillation. The $^{\n{nat}}$Kr~concentration in Xe has been lowered to $(19 \pm 4)$\,ppt, as measured in a Xe~gas sample from the detector using ultra-sensitive rare gas mass spec\-tro\-metry combined with a sophisticated Kr/Xe separation technique. This is consistent with the $(18 \pm 8)$\,ppt derived from counting the number of delayed $\beta$-$\gamma$ coincidences associated with the $^{85}$Kr beta decay, assuming a $^{85}$Kr/$^{\n{nat}}$Kr ratio of $2\times 10^{-11}$~\cite{ref::krcon}.

The data have been acquired under improved electronic noise conditions and with a modified trigger logic allowing to trigger on $>99\%$ of events with an S2 above 150 photoelectrons (PE). This has been directly measured using the method described in~\cite{Aprile:2012instr,ref::xe100analysisPaper} and leads to virtually no loss of events in the energy region of interest. 

The non-uniform light collection by the two PMT arrays and the attenuation of the ionization signal by residual impurities over the maximum drift gap of 30\,cm lead to a position-dependent S1 and S2~signal response. The signals are corrected using maps derived from calibration data.  
The S1~light yield is 3-dimensionally corrected in cylindrical coordinates $(r, \theta, z)$ in order to optimize the response very close to the PMTs.  The electron lifetime $\tau_e$~\cite{Aprile:2012instr}, used to describe the ionization loss by impurities in LXe, was measured regularly with a $^{137}$Cs source throughout the data taking period. The value increased from 374\,$\mu$s to 611\,$\mu$s, with the average being $\tau_e = 514$\,$\mu$s. The measured drift time $t_d$ is used to correct the S2~signal size for these losses, and an additional correction in ($x,y$) accounts for variations due to photon collection efficiency and small inhomogeneities in the mesh electrodes. The width of the S2 signal is also corrected in ($x,y,t_d$) such that it is independent of these inhomogeneities. For the analysis presented here, the maximum size of the latter two corrections is 15\% and 3\%, respectively. The corrections, including one due to the imperfect drift field, are described in more detail in~\cite{Aprile:2012instr}. The considerably larger amount of ER and NR calibration data taken during this dark matter run (48.0 and 2.7\,live days, respectively), 
allowed for the improvement of the accuracy of most of these corrections to the percent-level. 

As in the previous analysis, it was decided a priori to use the Profile Likelihood (PL) statistical inference as introduced in~\cite{Aprile:2011hx}. Both the signal and the background-only hypothesis are tested. An analysis based on the maximum gap method~\cite{Yellin:2002xd} with a pre-defined signal region (benchmark region), is used as a cross check.

The NR energy scale, E$_{\n{nr}}$, is derived from the S1 signal using the independently measured relative scintillation efficiency $\mathcal{L}_{\text{eff}}$ via the relation $E_{\n{nr}} \! = \! (S1/L_y) (1/\mathcal{L}_{\text{eff}})(S_{\n{ee}}/S_{\n{nr}})$ (see~\cite{Plante:2011hw} and references therein). The $\mathcal{L}_{\text{eff}}$ parametrization of \cite{Aprile:2011run08}, based on all the available direct measurements, is used. 
The factors $S_{\n{ee}}=0.58$ and $S_{\n{nr}}=0.95$ describe the scintillation quenching due to the electric field and are taken from~\cite{Aprile:2006kx}. 
$L_y=(2.28 \pm 0.04)$\,PE/keV$_{\n{ee}}$ is the updated response to 122~keV gamma rays as determined from calibration measurements using lines above and below this value. The interpolation between these lines is performed using the  Noble Element Simulation Technique (NEST) model for scintillation~\cite{Szydagis:2011tk}. 

After verification that electronic noise was not responsible for any of the S1~pulses, the lower energy threshold used for this analysis was set to 3\,PE, corresponding to 6.6\,keV$_{\n{nr}}$. The PL analysis takes into account the expected WIMP energy distribution and would not need an upper energy threshold. However, an upper threshold of 43.3\,keV$_{\n{nr}}$ (30\,PE) was employed and the data above this energy were used to test the background prediction before unblinding. The benchmark region is limited to an upper threshold of 30.5\,keV$_{\n{nr}}$ (20\,PE) chosen to optimize the signal-to-background ratio. Signal (NR) and background (ER) events can be distinguished by their different S2/S1 ratio, where only the S2 signal detected by the bottom PMTs, S2$_b$, is used since it requires smaller corrections~\cite{Aprile:2012instr}. The mean of the $\log_{10}$(S2$_b$/S1) band from ER calibration data is subtracted in order to remove the energy-dependence of this discrimination parameter.

The dark matter data used for this analysis were accumulated over a period of 13 months between February 28, 2011 and March 31, 2012. Besides 3\,interruptions due to equipment maintenance, the data were acquired continuously. 
Dark matter data taking was otherwise only interrupted by regular calibrations using blue LED light (for the PMT response), a $^{137}$Cs source (for monitoring of the LXe purity), and $^{60}$Co and $^{232}$Th sources (for ER background calibration). Overall, the duty cycle of XENON100 during this dark matter run was 81\%. To calibrate the response to NRs, 
data from an $^{241}$AmBe neutron source were taken just before the start and right after the end of the run. 
The two measurements are in good agreement. 

Periods with increased electronic noise or very localized light emission in the $xy$-plane were removed from the data, as well as periods in which crucial detector parameters such as temperature or pressure fluctuated outside of their normal range (3\% of the total data taking time). This results in a final dark matter dataset of 224.6\,live days. In order to avoid analysis bias, the dark matter data were blinded from 2-100\,PE in S1 by keeping only the upper 90\% of the ER band, thereby masking more than 90\% of the signal region.

In order to identify valid NR candidate events with the highest possible acceptance, several classes of cuts and event selections are applied to the data. Their acceptance is evaluated on NR calibration data, with the exception of quality cuts which might have a time dependence due to changing detector conditions. These are tested on the non-blind part of the science data or on the ER calibration data. The first class of cuts are basic data quality cuts which remove events that show either unidentified peaks or an excessive level of electronic noise or light (as from a very high energy event or a high voltage discharge). Since only single-scatter events are expected from WIMP interactions, the second class of cuts identifies such events using the number of S1 and S2 peaks in the waveform and the information from the active veto. 
Conditions on the size of the S2 and the requirement that at least two PMTs must observe an S1 peak ensure that only data above the threshold and well above the noise level are considered. Finally, it is verified that several quantities associated with the event are consistent, e.g., that the width of the S2 signal, affected by electron diffusion in the LXe, is consistent with the $z$-position derived from the time difference between the S1 and the S2. 

\begin{figure}[b]
\centering
\includegraphics[width=1\columnwidth]{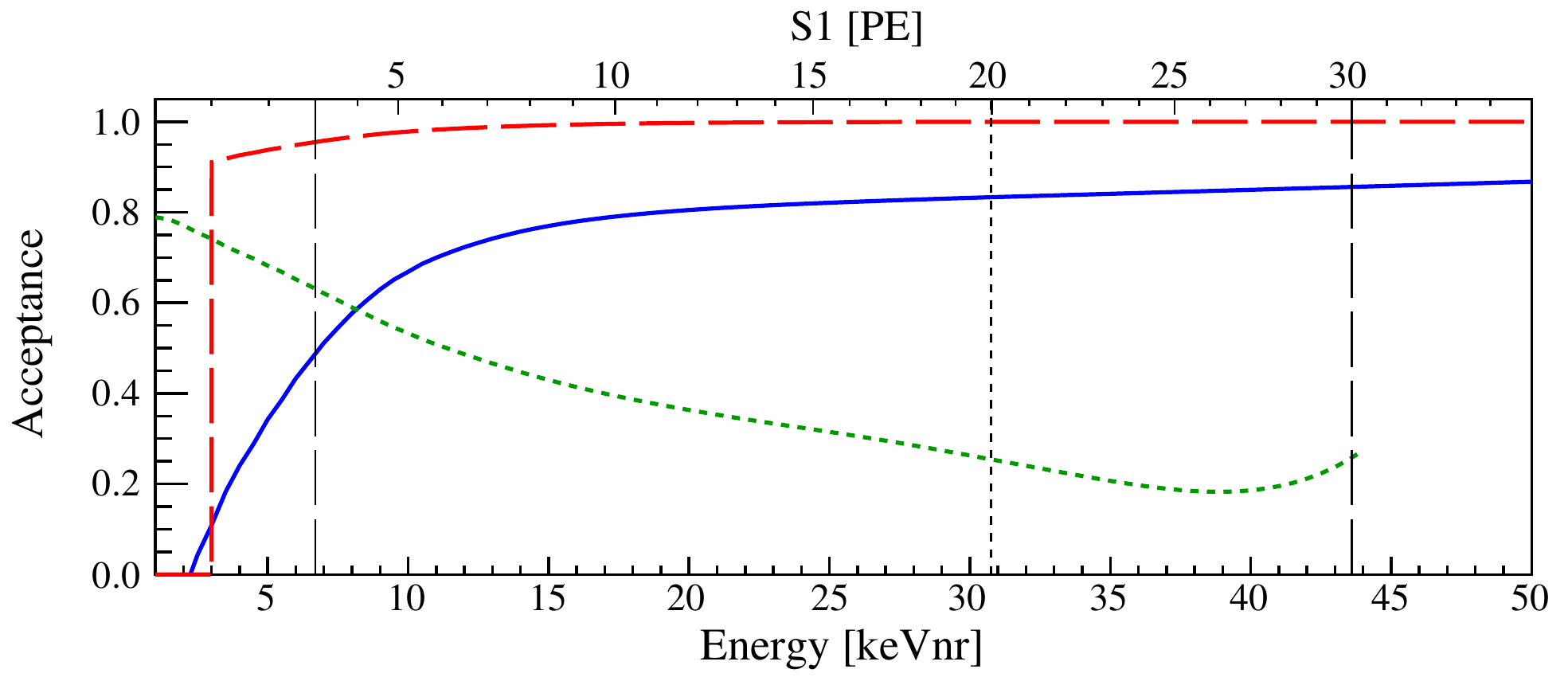}
\caption{Combined cut acceptance (solid blue). The S2 threshold cut S2$>$150\,PE (dashed red) is independent of possible fluctuations in S1 and has to be applied to the S1 spectrum before taking into account the S1 resolution. It is conservatively set to zero below 1\,PE. For the cross-check with the maximum gap method~\cite{Yellin:2002xd}, a hard discrimination cut is used. Its acceptance to NRs is shown by the dotted green line. The lower analysis threshold is 6.6\,keV$_{\n{nr}}$ (3\,PE) and extends to 43.3\,keV$_{\n{nr}}$, whereas the cross-check is restricted to 30.5\,keV$_{\n{nr}}$ (dashed and dotted black lines).}\label{fig:acceptance}
\end{figure}

The cuts and their acceptance determination are identical or similar to those described in~\cite{ref::xe100analysisPaper} except for the cut against electronic noise which has been improved considerably. The acceptance of most of the cuts is given vs. the size of the measured S1 signal, used to infer the NR energy scale (blue line in Fig.~\ref{fig:acceptance}). The S1 signal is subject to large Poisson fluctuations due to the low number of quanta involved. Only the acceptance of the S2 threshold condition, S2$>$150\,PE, is given vs.~the S1 signal before Poisson-smearing since the S2 signal fluctuates independently  from the S1 after the initial energy deposition. Given the systematic uncertainties in the LXe light and charge yields and the resulting XENON100 response at very low nuclear recoil energies, we choose not to model WIMP interactions with energy deposits below 3\,keV$_{\n{nr}}$. With the mean $\mathcal{L}_{\text{eff}}$ shown in~\cite{Aprile:2011run08}, this is esentially equivalent to neglecting upward fluctuations in S1 above the threshold of 3\,PE from energy deposits with S1 expectation values below 1 PE. This approach results in a conservative upper limit for low mass WIMPs. For the central part of the WIMP mass range its impact is  $<1\%$. The resulting acceptance of the S2$>$150 PE cut~\cite{ref::xe100analysisPaper} is shown in Fig.\,\ref{fig:acceptance} (red line).

\begin{figure}[b]
\centering
\includegraphics[width=1\columnwidth]{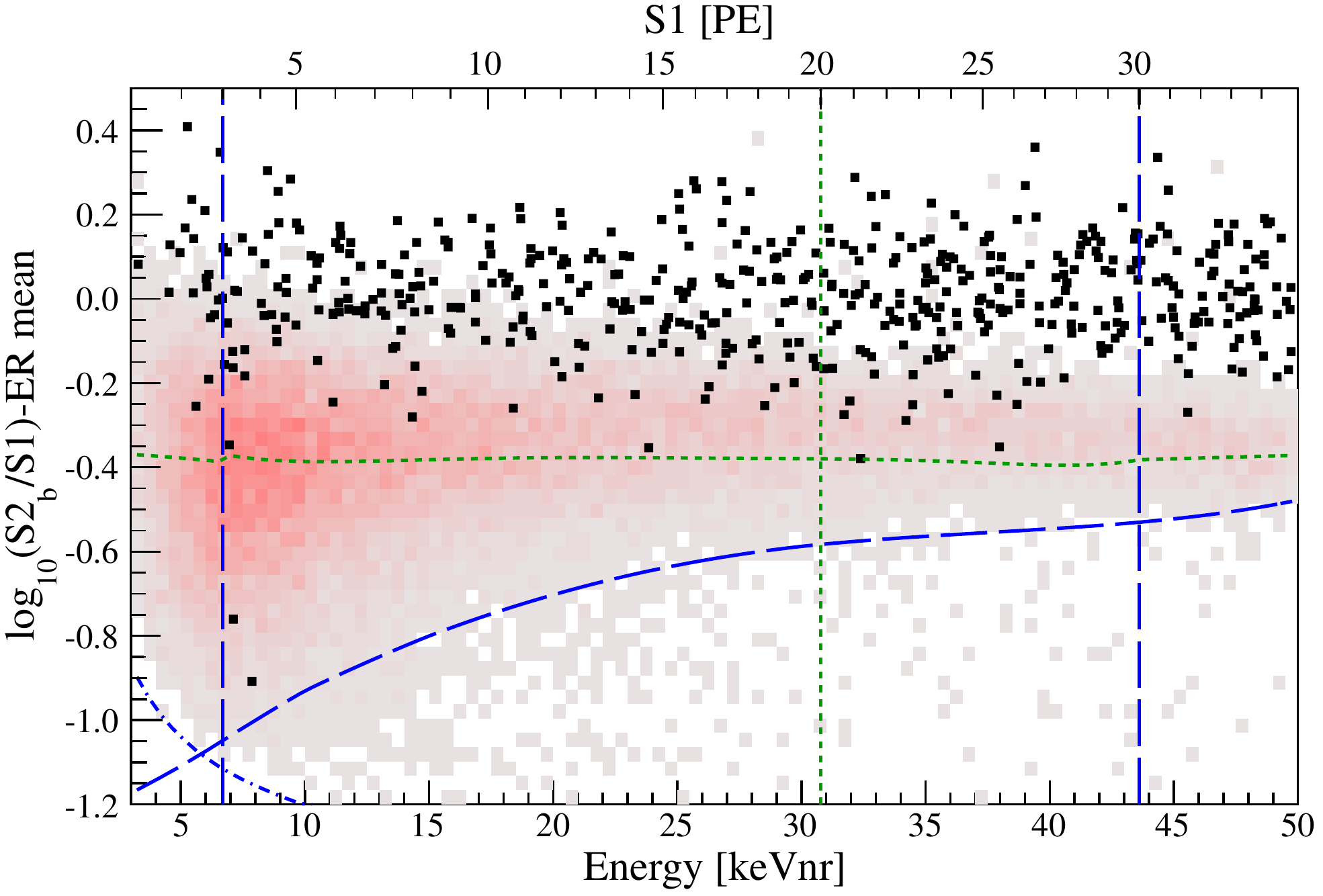}
\vspace{0.08cm}\footnotesize{ \ }
\includegraphics[width=1\columnwidth]{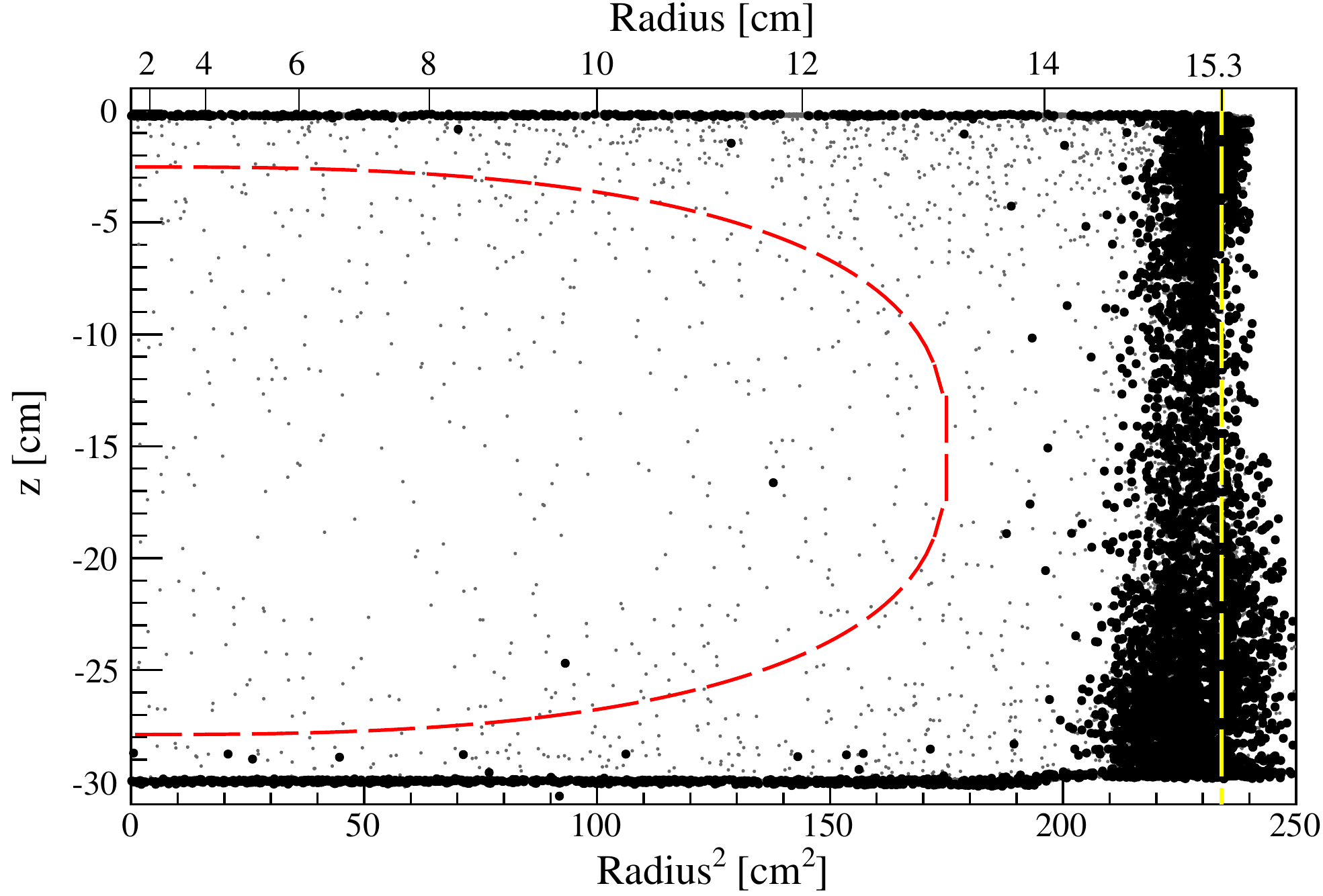}
\caption{(Top) Event distribution in the discrimination parameter space $\log_{10}$(S2$_b$/S1), flattened by subtracting the distribution's mean, as observed after unblinding using all analysis cuts and a 34~kg fiducial volume (black squares). A lower analysis threshold of 6.6\,keV$_{\n{nr}}$ (NR equivalent energy scale) is employed. The PL analysis uses an upper energy threshold of 43.3~keV$_{\n{nr}}$ (3-30~PE) and the benchmark WIMP search region is limited to 30.5\,keV$_{\n{nr}}$ (3-20\,PE). The negligible impact of the S2$>$150\,PE threshold cut is indicated by the dashed-dotted blue line and the signal region is restricted by a lower border running along the $97\%$ NR quantile. An additional hard S2$_b$/S1 discrimination cut at 99.75\% ER rejection defines the benchmark WIMP search region from above (dotted green) but is only used to cross check the PL inference. The histogram in red/gray indicates the NR band from the neutron calibration. Two events fall into the benchmark region where $(1.0 \pm 0.2)$ are expected from background. 
(Bottom) Spatial event distribution inside the TPC using a 6.6-43.3\,keV$_{\n{nr}}$ energy window. The 34\,kg fiducial volume is indicated by the red dashed line. Gray points are above the 99.75\% rejection line, black circles fall below.}\label{fig::events}
\end{figure}

The fiducial volume used in this analysis contains 34\,kg of LXe. The volume was determined before the unblinding by maximizing the dark matter sensitivity of the data given the accessible ER background above the blinding cut. The ellipsoidal shape was optimized on ER calibration data, also taking into account event leakage into the signal region. A benchmark WIMP search region to quantify the background expectation and to be used for the maximum gap analysis was defined from 6.6-30.5\,keV$_{\n{nr}}$ (3-20\,PE) in energy, by an upper 99.75\%~ER rejection line in the discrimination parameter space, and by the lines corresponding to S2$>$150~PE and a lower line at $\sim$97\% acceptance from neutron calibration data (see lines in Fig.\,\ref{fig::events}, top).

Both NR and ER interactions contribute to the expected background for the WIMP search. The first is determined from Monte Carlo simulations, using the measured intrinsic radioactive contamination of all detector and shield materials~\cite{Aprile:2011ru} to calculate the neutron background from ($\alpha,n$) and spontaneous fission reactions, as well as from muons, taking into account the muon energy and angular dependence at LNGS. The expectation from these neutron sources is $(0.17 {+0.12 \atop -0.07})$\,events for the given exposure and NR acceptance in the benchmark region. About 70\%~of the neutron background is muon-induced. 

ER background events originate from radioactivity of the detector components and from $\beta$ and $\gamma$ activity of intrinsic radioactivity in the LXe target, such as $^{222}$Rn and $^{85}$Kr. The latter background is most critical since it cannot be reduced by fiducialization. 
Hence, for the dark matter search reported here, a major effort was made to reduce the $^{85}$Kr contamination which affected the sensitivity of the previous search~\cite{Aprile:2011run08}.
To estimate the total ER background from all sources, the $^{60}$Co and $^{232}$Th calibration data is used, with $>$35\,times more statistics in the relevant energy range than in the dark matter data. 
The calibration data is scaled to the dark matter exposure by normalizing it to the number of events seen above the blinding cut in the energy region of interest. The majority of ER background events is Gaussian distributed in the discrimination parameter space, with a few events leaking anomalously into the NR band. 
These anomalous events can be due to double scatters with one energy deposition inside the TPC and another one in a charge insensitive region, such that the prompt S1 signal from the two scatters is combined with only one charge signal S2. Following the observed distribution in the calibration data, the anomalous leakage events were parametrized by a constant (exponential) function in the discrimination parameter (S1 space).
The ER background estimate including Gaussian and anomalous events is $(0.79 \pm 0.16)$ in the benchmark region, leading to a total background expectation of $(1.0 \pm 0.2)$\,events. 

The background model used in the PL analysis employs the same assumptions and input spectra from MC and calibration data. Its validity has been confirmed prior to unblinding on the high-energy sideband and on the vetoed data from 6.6-43.3\,keV$_{\n{nr}}$.

After unblinding, two events were observed in the benchmark WIMP search region, see~Fig.\,\ref{fig::events}. With energies of 7.1\,keV$_{\n{nr}}$ (3.3\,PE) and 7.8\,keV$_{\n{nr}}$ (3.8\,PE) both fall into the lowest PE bin used for this analysis. The waveforms for both events are of high quality and their S2/S1\,value is at the lower edge of the NR band from neutron calibration. There are no leakage events below 3~PE.
The PL analysis yields a $p$-value of $\ge5$\% for all WIMP masses for the background-only hypothesis indicating that there is no excess due to a dark matter signal. The probability that the expected background in the benchmark region fluctuates to 2\,events is 26.4\% and confirms this conclusion.

\begin{figure}[t]
\centering
\includegraphics[width=1\columnwidth]{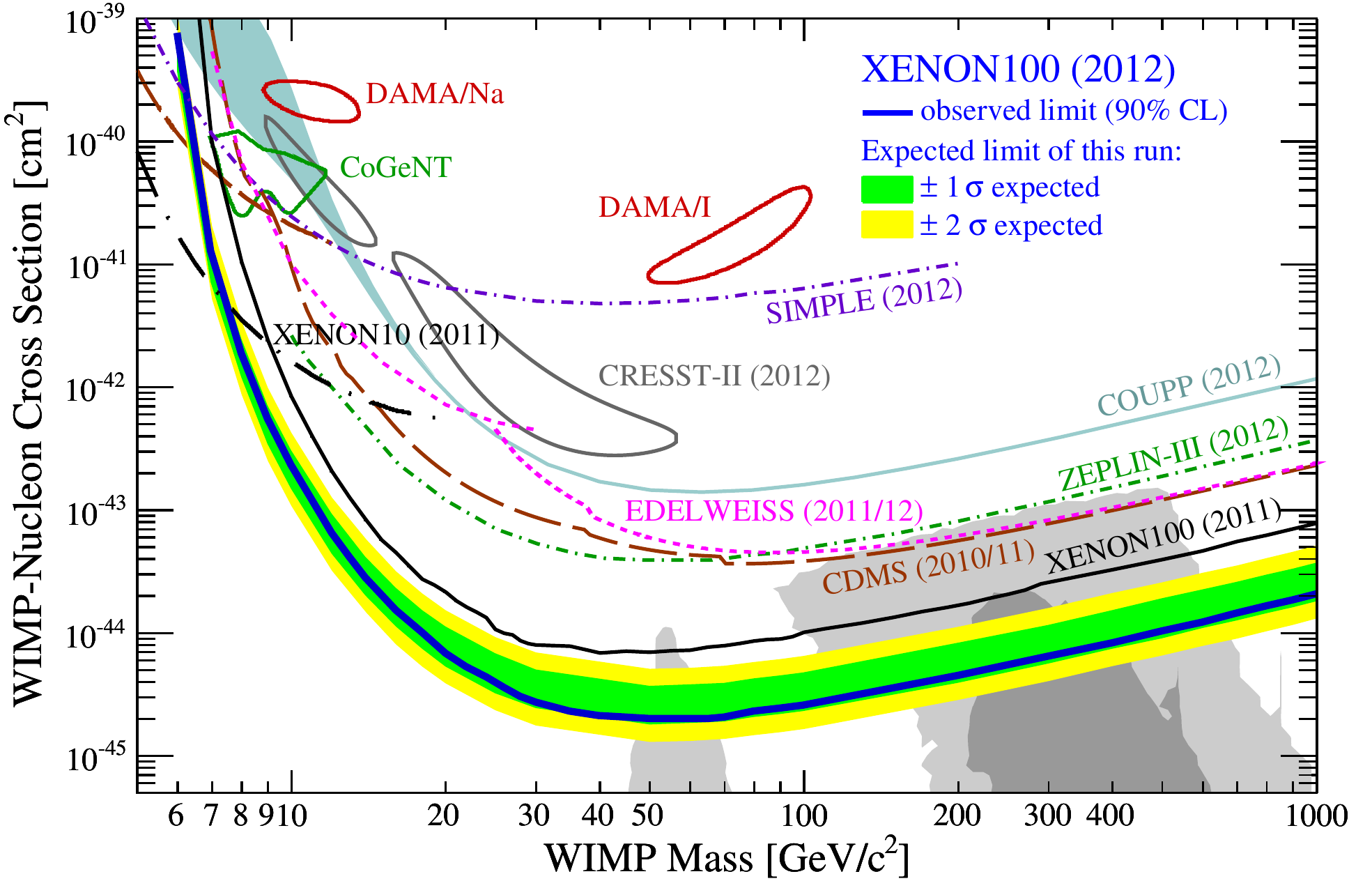}
\caption{Result on spin-independent WIMP-nucleon scattering from XENON100: The expected sensitivity of this run is shown by the green/yellow band ($1\sigma/2\sigma)$ and the resulting exclusion limit (90\% CL) in blue. 
For comparison, other experimental limits (90\% CL) and detection claims ($2\sigma$) are also shown~\cite{Savage:2008er,Aalseth:2010vx,Angloher:2011uu,Ahmed:2009zw}, 
together with the regions ($1\sigma$/$2\sigma$) preferred by supersymmetric (CMSSM) models~\cite{ref::susymodels}.}\label{fig::limit}
\end{figure}

A 90\% confidence level exclusion limit for spin-independent WIMP-nucleon cross sections $\sigma_\chi$ is calculated, assuming an isothermal WIMP halo with a local density of $\rho_\chi=0.3$\,GeV/$cm^3$, a local circular velocity of $v_0=220$\,km/s, and a Galactic escape velocity of $v_{\mathrm{esc}}=544$\,km/s~\cite{smith:rave}. 
Systematic uncertainties in the energy scale as described by the $\mathcal{L}_{\text{eff}}$ parametrization of~\cite{Aprile:2011run08} and in the background expectation are profiled out and represented in the limit. Poisson fluctuations in the number of PEs dominate the S1~energy resolution and are also taken into account along with  the single PE resolution. The expected sensitivity of this dataset in absence of any signal is shown by the green/yellow ($1\sigma/2\sigma)$ band in Fig.~\ref{fig::limit}. The new limit is represented by the thick blue line. It excludes a large fraction of previously unexplored parameter space, including regions preferred by scans of the constrained supersymmetric parameter space~\cite{ref::susymodels}. 

The new XENON100 data provide the most stringent limit for $m_\chi>8$\,GeV/$c^2$ with a minimum of $\sigma=2.0 \times 10^{-45}$\,cm$^2$ at $m_\chi=55$\,GeV/$c^2$. The maximum gap analysis uses an acceptance-corrected exposure of 2323.7\,kg$\times$days (weighted with the spectrum of a 100\,GeV/$c^2$ WIMP) and yields a result which agrees with the result of Fig.\,\ref{fig::limit} within the known systematic differences. The new XENON100 result continues to challenge the interpretation of the DAMA~\cite{Savage:2008er}, CoGeNT~\cite{Aalseth:2010vx}, and CRESST-II~\cite{Angloher:2011uu} results as being due to scalar WIMP-nucleon interactions.

We acknowledge support from NSF, DOE, SNF, UZH, Volkswagen Foundation, FCT, R\'egion des Pays de la Loire, STCSM, NSFC, DFG, Stichting FOM, Weizmann Institute of Science, and the friends of Weizmann Institute in memory of Richard Kronstein. We are grateful to LNGS for hosting and supporting XENON.

\newpage
{ \ }
\newpage
{ \ }

\includepdf{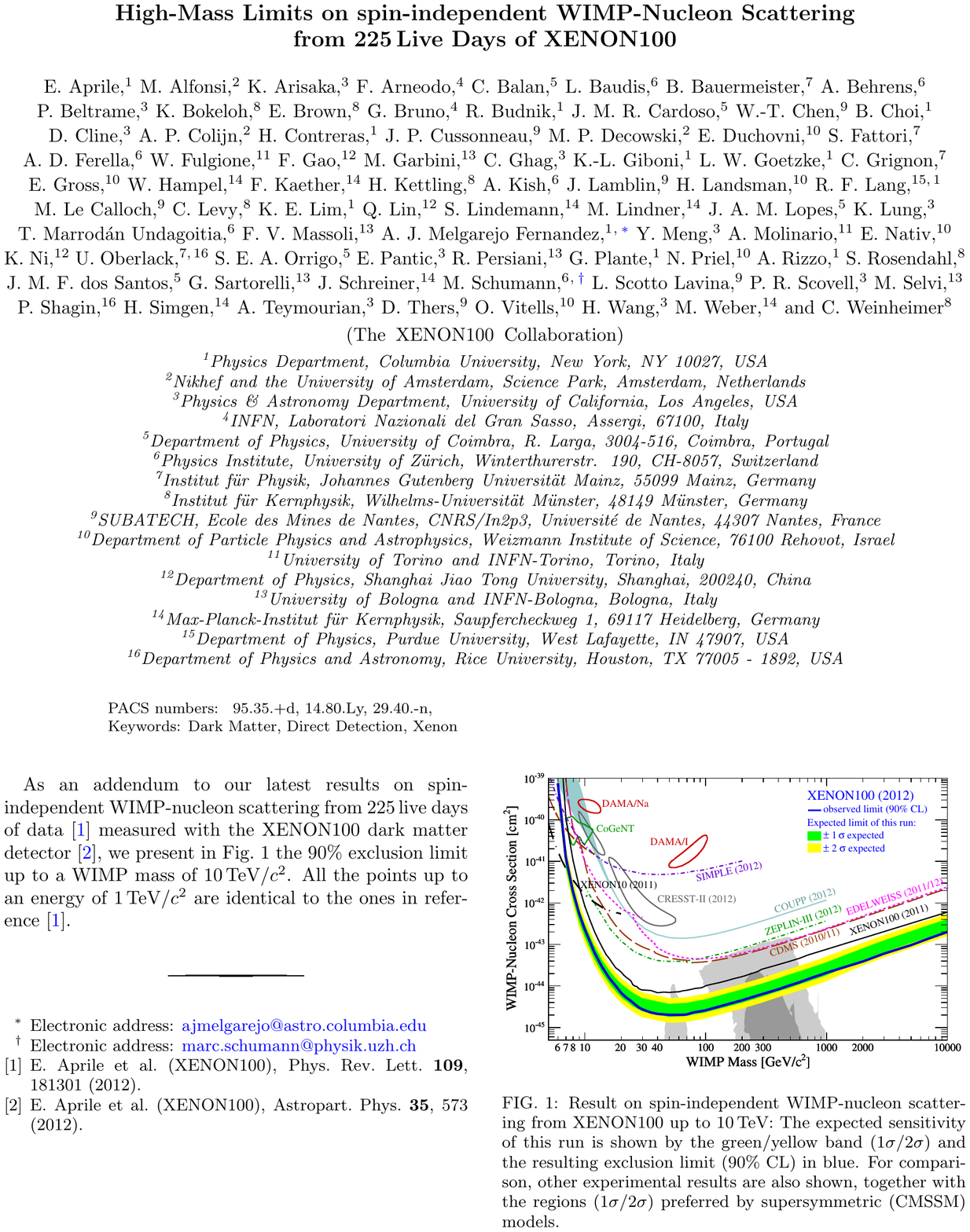}

\end{document}